\documentclass[12pt,preprint]{aastex}

\newcommand{\LCDM}{$\Lambda$CDM\ }

\newcommand{\kms}{\hbox{\rm km\,s$^{-1}$}\,}

\newcommand{\etal}{{ et~al.~}}

\newcommand{\Ms}{M_\odot}

\begin{document}


\title{Numerical Simulations of High Redshift Star Formation in Dwarf
Galaxies}

\author{Konstantinos Tassis}
\affil{Department of Astronomy, University of Illinois at
Urbana-Champaign, \\ 1002 W. Green Street, Urbana, IL, US--61801}
\email{tassis@astro.uiuc.edu}

\author{Tom Abel}
\affil{Penn State University, University Park, PA, US--16802}

\author{Greg L. Bryan}
\affil{University of Oxford, Astrophysics, Keble Road, Oxford OX1
3RH, UK}

\author{Michael L. Norman}
\affil{University of California, San Diego, CA, US--92093 La Jolla}


\begin{abstract}
  
We present first results from three-dimensional hydrodynamic
simulations of the high redshift formation of dwarf galaxies.  The
simulations use an Eulerian adaptive mesh refinement technique to
follow the non--equilibrium chemistry of hydrogen and helium with
cosmological initial conditions drawn from a popular
$\Lambda$-dominated cold dark matter model.  We include the effects of
reionization using a uniform radiation field, a phenomenological
description of the effect of star formation and, in a separate
simulation, the effects of stellar feedback. The results highlight the
effects of stellar feedback and photoionization on the baryon content
and star formation of galaxies with virial temperatures of
approximately $10^4$K.  Dwarf sized dark matter halos that assemble
prior to reionization are able to form stars. Most halos of similar
mass that assemble after reionization do not form stars by redshift of
three.  Dwarf galaxies that form stars show large variations in their
gas content because of stellar feedback and photoionization effects.
Baryon-to-dark matter mass ratios are found to lie below the cosmic
mean as a result of stellar feedback.  The supposed substructure
problem of CDM is critically assessed on the basis of these results.
The star formation histories modulated by radiative and stellar
feedbacks are discussed. In addition, metallicities of individual
objects are shown to be naturally correlated with their mass-to-light
ratios as is also evident in the properties of local dwarf galaxies.
\end{abstract}

\keywords{cosmology: theory -- early universe -- galaxies: formation }


\section{Introduction}
\label{sec:introduction}

Radiative and stellar feedback have long been recognized as an
essential ingredient in galaxy formation and they are thought to 
play an important role in determining the properties of 
dwarf galaxies. 
The smallest dark matter halos are thought to be the sites for the
formation of the first stars (Couchman and Rees 1986; Tegmark et
al. 1997, Haiman and Loeb 1997; Abel et al. 1998; Bromm, Coppi and
Larson 1999; Abel, Bryan and Norman 2000, 2001). Whether these
smallest objects will allow a significant fraction of the baryons to
participate in primordial star formation is unclear because of the
effect of complex radiative feedback processes on molecular hydrogen
formation (Haiman, Abel and Rees 2000; Ciardi, Ferrara, Abel 2000;
Mahacek, Bryan and Abel 2001, Glover and Brand 2001).  However, the
early formation of even a few stars in such low-mass objects may
significantly affect the evolution of their baryon content due to
supernova explosions driving a significant fraction of the gas away
from the gravitational potential of the host.


Stellar feedback was first discussed in the context of primordial
globular cluster formation by Peebles and Dicke (1968).  In the
context of hierarchical structure formation, the possibility that
supernovae may remove a significant fraction of the gas present in
dwarf galaxies was discussed by Dekel \& Silk (1986) and by Couchman
\& Rees (1986). Further analytical treatment of such supernova-driven
winds was presented by Babul \& Rees (1992) and Efstathiou (1992).

Star formation and feedback have also been studied in the 
context of cosmological structure formation simulations. 
However, the efficiency of supernova energy feedback in 
driving the gas out of the dark matter potentials was 
found to depend on the details of the feedback prescription: 
Katz (1992) and Weinberg, Hernquist \& Katz (1997) found that 
the properties of their simulated galaxies were insensitive to 
the inclusion of star formation and feedback to the simulations. 
Navarro \& White (1993) and Navarro \& Steinmetz (1999, 2000) 
found that the effect of supernova feedback depends on whether the 
energy returned to the ISM is in the form of thermal or kinetic
energy. In the former case, the energy was quickly radiated 
away and the effect of feedback was minimal, while
in the latter case feedback was efficient in driving gas away 
from the dark matter potentials and in suppressing star 
formation. In addition, Mac Low \& Ferrara (1999) and Navarro 
\& Steinmetz (2000) found that the effect of feedback depends on the 
total mass of the system under consideration, with the low-mass 
systems being affected much more than the higher-mass systems. 

Energy feedback from supernovae is also believed to have a significant
effect on the metallicities of dwarf galaxies. The idea of metals
being driven away from galaxies is supported by observations of the
chemical enrichment of the IGM (e.g. Ellison et al. 2001 and
references therein) as well as observations of dwarf galaxies in he
Local Group.  Prada \& Burkert (2002) have summarized a correlation
between mass-to-light ratio and metallicity of dwarf spheroidals,
which they attribute to metal loss being more efficient for objects
with shallower potential wells.

The ability of supernova-driven winds to remove metals from galaxies has been
studied by Mac Low \& Ferrara (1999) and by Aguirre \etal (2001). 
Mac Low \& Ferrara (1999) modeled the effects of repeated 
supernova explosions from starbursts in dwarf galaxies with hydrodynamic
simulations performed in a pre-determined dark matter gravitational 
potential and with an initial baryon-to-gas mass ratio extrapolated from
observations of higher mass systems. Their results suggest that galactic 
winds are very efficient in removing metals from dwarf galaxies, with the 
effect being more pronounced for the least massive objects.
For larger energy ejection rates (Mac Low \& Ferrera considered star
formation rates only up to about 0.004 $M_\odot$/year), the gas in the
galaxy can be completely blown away (Mori, Ferrara \& Madau 2002).

Aguirre \etal (2001) used results from completed N-body + 
hydrodynamic simulations including star formation which, however, did
not directly incorporate the result of feedback, and 
assumed a series of  ad hoc algorithms
to reconstruct the metal enrichment process starting from the star
formation history in each simulated object. They found that 
most galaxies at redshift $\sim 3$ should exhibit winds
driving metals away from their gravitational potentials and 
enriching the IGM with metals. Note that such winds can now be studied 
directly using aborption spectra of quasars behind Lyman break galaxies 
(Adelberger et al. 2002).


An additional factor affecting the baryon contents of dwarf galaxies
is photoionization.  The Jeans mass prior to reionization is
approximately a factor $10^5$ smaller than after,  
i.e.  the minimum masses of
dark matter halos that can confine baryons are much less at times
before the intergalactic medium has been ionized.

Efstathiou (1992) and Babul \& Rees (1992) discussed analytically the
possibility that a photoionizing background might significantly affect
the formation of dwarf galaxies and may be the mechanism preventing
the collapse of most of the baryonic material into subgalactic objects
at high redshifts.  Thoul and Weinberg (1996) showed, using a
one--dimensional hydrodynamics code, that photo-ionization affects the
collapse of gas into dark halos with velocity dispersions even in
excess of $30\kms$.

Kepner, Babul and Spergel (1997) have
looked at the formation of molecular hydrogen in the cores of
proto--dwarf galaxies and Kepner et al. (1999) 
investigated observational consequences of
the picture of delayed dwarf galaxy formation. The scenario for
the formation of dwarf galaxies outlined by Ferrara and Tolstoy (2000)
also takes into account the feedback from photo-ionization. 
In their model the first stars in observed dwarfs form prior to 
reionization. Star formation at later epochs originates from new gas 
infall as the meta--galactic UV background radiation
field decreases.  Gnedin (2000) studied the effect of photoionization on the 
gas fraction of low-mass objects and found a deviation from the 
universal baryon fraction at low masses.

Many of these papers, as well as Ciardi et
al. (2000), did implicitly predict that not all the substructure 
seen in dark matter--only simulations should also host luminous objects. 
Images of multiply lensed quasars may be used to constrain the amount 
of dark substructure in galaxies
(Metcalf and Madau 2001). Interestingly, existing data on such systems
favors the existence of such dark matter substructure (Metcalf and Zhao 
2002, Dalal and Kochanek 2002).

In this paper, we use high-mass resolution, 3-dimensional hydrodynamic
AMR simulations to examine the effect of stellar feedback and
photoionization on the gas content, star formation histories and
metallicities of dwarf galaxies down to a redshift of 3.  Star
formation and feedback are included using a phenomenological
recipe. Since star formation, metal and energy feedback are followed
concurrently with the dark matter and gas evolution, our simulations
{\em do not require invoking any assumptions } on the nature of
supernova driven winds (as in Aguirre \etal (2001)) or on the shape of
the dark matter gravitational potentials (as in Mac Low \& Ferrara
(1999)).  In addition, grid based methods are known to be superior to 
particle techniques when a multi-phase medium needs to be resolved (Croft 
et al. 2001, Pearce et al. 2000, Ritchie \& Thomas 2001, and references
therein). Here, we will only discuss aspects of the simulation that
are expected to be insensitive to the details of the star formation
algorithm.

Our paper is organized as follows: 
In \S 2 we describe the simulations carried out as well as relevant 
aspects of our implementation of star formation and feedback. In \S 3 we 
discuss the global star formation history in our cube before and after 
reionization. In \S 4 we present the effects of feedback and reionization 
on the baryon contents, star formation histories and metallicities of 
individual halos.  Finally, we discuss our findings in \S 5.


\section{Simulations}

We have used the 3-D AMR cosmology code described by Bryan \& Norman
(1997a), Norman and Bryan (1999), Bryan, Abel and Norman (2001).  The AMR
algorithm used is similar to the one described by Berger and Collela 
(1989).  It utilizes an adaptive hierarchy of grid patches at various 
levels of resolution. Each rectangular grid patch covers the region of 
space in its parent grid needing higher resolution, and may itself become 
the parent grid to an even higher resolution child grid. Initial and 
boundary conditions are obtained by interpolation from the parent grid 
when the child grid is created.

In the simulations presented in this paper, we limited the refinement 
to 9 levels in a $128^3$ top grid. However, at the end of each run, only 7 
levels of refinement had been used, so our simulations were not resource 
bounded. We have used a box size of $10\, h^{-1}$ Mpc comoving, which was  
chosen as a compromise between high mass resolution and reasonable volume.
The maximum spatial resolution (i.e. minimum cell size) achieved was 
610$h^{-1}$  comoving pc or better than 150 proper pc for the redshifts 
simulated.  The dark matter mass 
resolution was $M_{\rm DM} \approx 5 \times 10^7 M_\odot$.

While this resolution is insufficient to model the detailed internal
structure of the galaxies and the way that feedback works on pc scales, it
is enough to correctly model the conversion of SN energy into
high-velocity outflows which affect the gas surrounding (and infalling
into) galaxies.  For a more detailed examination of the impact of a
burst of SN within a single high-redshift halo of this mass, see Mori,
Ferrara \& Madau (2002); their results support our findings that massive
outflows are common around starbursting galaxies.

The simulations were set up as a flat \LCDM model with
$\Omega_\Lambda=0.7$ and $\Omega_m=0.3$. The total matter density 
consists 
of a baryonic and a dark matter component, where $\Omega_{b}=0.04$ and 
$\Omega_{DM}=0.26$.  The present-day value of the dimensionless Hubble 
parameter was $h=0.67$ in units of $100 \,{\rm km/s/Mpc}$.  The 
simulations were initialized at redshift 60 and terminated at redshift 3. 
The cosmological density field was initialized using
power spectra for the dark matter and the baryons given by the 
fitting functions of Eisenstein \& Hu (1999).

Our simulations have allowed for star formation, however 
radiation feedback from the formed ``stars'' was not taken 
into account - instead, the simulation implements a homogeneous radiation 
field of Haardt \& Madau type (Haardt and Madau 1996), with an index of 
$-1.5$, i.e. full reionization occurs in our models at $z\sim 6$. 
\footnote{A more 
refined study of cosmological reionization using the galaxies formed in 
our no--feedback run has been presented in Razoumov et al (2002).} We also 
solve the non-equilibrium chemistry and cooling from hydrogen and helium 
and their ions (Abel et al. 1997, Anninos et al.  1997). 
Cooling from metal lines has been approximated by assuming collisional 
equlibrium. However, it should be noted that in the case of galaxy 
formation radiative processes may strongly affect the ionization
balances and hence the cooling rates.

In order to allow for star formation, given that 
it is not feasible to directly simulate the formation
of individual stars since the range of scales and additional physics
required to do so is prohibitive,  a parametric method was 
adopted which attempts to model star-forming regions following a 
prescription similar to the one introduced by Cen and Ostriker (1992). A 
detailed description of the implementation of the star-formation and 
feedback scheme can be found in O'Shea \etal (2002).

Groups of stars forming are treated as pressureless particles which
return feedback to the surrounding medium (star particles).  For a star
particle to be created, the following criteria need to be met:
\begin{enumerate}
\item The baryon density at the particular grid where the star
particle is to be created exceeds the mean density in the universe 
by a predefined factor, set to be equal to 100 in our simulation.
\item The gas in the particular grid cell has $\nabla \cdot \mbox{\bf v} <
0$, i.e. the gas is infalling rather than expanding (or $T < 10^4$ K).
\item The cooling timescale for the gas is smaller than the dynamical
collapse timescale.
\item The baryonic mass in the cell exceeds the local Jeans mass.
\end{enumerate}
The mass of the star particle to be created is calculated as
\begin{equation}\label{mstar}
m_*=f_{\rm *eff} \, m_b \frac{\Delta t}{t_{\rm dyn}} 
\end{equation}
where $m_b$ is the baryon mass of the cell, $\Delta t$ is the duration
of a timestep, $t_{\rm dyn}$ is the dynamical collapse timescale and
$f_{\rm *eff}$ is the ``star formation efficiency'' (fraction of mass of 
the initial cloud which is converted into stars), set equal to 80\% in 
this simulation. This value of $f_{\rm *eff}$ is significantly larger 
than values expected in realistic systems 
but 
we have used it in order to investigate the limiting 
case as to the maximal effect of feedback. Additionally, we never allow 
more than $90\%$ of the mass in a cell to be 
converted to stars. 

Conditions 1-4 are necessary but not sufficient for a star particle to
be formed: in order to prevent the formation of an excessive number of
star particles which would require a prohibitively high amount of
computational resources in order to be followed, a minimum star particle 
mass $m_{\rm *min}$ is introduced.  A star particle is formed in a cell 
fulfilling the criteria 1-4 only if $m_*$ exceeds this minimum mass. The 
value of $m_{\rm *min}$ we used was $10^6 {\rm M_\odot}$.  However, since 
the minimum mass requirement is associated with computational resources 
limitations rather than underlying physics, a ``bypass mechanism'' for 
this restriction was included in the algorithm to account for small-mass 
star particles which should have been created on physical grounds: there 
is a certain probability per time step that a star particle with mass 
lower than the minimum mass threshold is created in a cell fulfilling the 
requirements 1-4 but not the minimum mass  requirement. This probability 
is equal to the ratio of the mass of the star particle that would be 
created in the cell over the minimum star particle  threshold.

To avoid the unphysical situation of forming a star particle
(corresponding to a cluster of stars) instantaneously, a more reasonable 
star formation history has been assigned to each cold collapsing cloud.  
Therefore, each star particle $i$ is assumed to have a 
time-dependent star 
formation rate of
\begin{equation} \label{ind_sfr}
\psi _i (t) = m_* \frac{t-t_{\rm cr}}{\tau^2}
\exp (-\frac{t-t_{\rm cr}}{\tau})
\end{equation}
where $t_{\rm cr}$ is the time of creation of the star particle and
$\tau = \max [t_{\rm dyn}, 10 {\rm Myrs}]$, with $t_{\rm dyn}$ being the
dynamical time from equation (\ref{mstar}).  Thus, one observes a rapid 
star formation after a dynamical time with an exponential tail at longer 
times.  If $t_{\rm dyn}$ is shorter than the small but non-zero life-span 
of massive  stars, an estimate of the latter (here taken to be 10 Myrs) is 
used in place of $t_{\rm dyn}$.

The star particles formed give energy, metals and momentum feedback to
the baryon gas. The energy equivalent to $10^{-5}$ of the rest mass energy
of the stars generated 
is converted into thermal energy 
(this corresponds to 1 SNe of $10^{51}$ erg for 
every 55M$_\odot$ of stars being formed) 
over a time period
corresponding to the dynamical time from equation (\ref{mstar}) and is
put directly into the thermal energy of the cell in which the star
particle resides.  Metals are also returned to the gas over the 
same time period and the yield (mass of metals generated per mass of stars 
created) is 0.02. The mass fraction of created 
stars which is returned to the gas phase is 25\%. The values for 
the parameters of our star formation algorithm are taken from 
Cen \& Ostriker (1992).

In order to better understand which effects in the output of our
simulation can be attributed to the feedback from the star particles to 
the baryonic gas, we have run a second, control simulation, with the exact 
same cosmological parameters and initial conditions as our original 
simulation, in which stellar particles are born following the exact same 
recipe as before, but they do not return energy, momentum or metals to the 
gas. In the following sections we will present and compare results from 
both simulations. We will refer to the original simulation (the one which 
allows for stellar feedback) as ``feedback run'' and to the control 
simulation (the one without stellar feedback) as ``no-feedback run''.
It is interesting to note that the two simulations presented here are
both limiting cases as far as feedback is concerned, in the sense that in
our original simulation the feedback from each individual star particle is
rather overestimated compared to theoretical expectations for realistic
physical systems, while in our second (control) simulation the feedback is
entirely absent.

\section{The Global Star Formation History}

Fig.~\ref{fig:gsfr} gives the global star formation rate in our
simulated box (in ${\rm M_\odot}$ per year per comoving ${\rm Mpc^3}$
) as a function of redshift. In the same plot, we have overplotted
observational data points compiled by Blain (2001). The observational
points are meant to give a feeling of the magnitude of the star
formation rate density in our box in terms of the corresponding
quantity observed in the real universe rather than serve as an
observational test for our model. The unrealistically large value of
the star formation efficiency, combined with the small box size, can
result to rapid gas depletion, causing the cosmic star formation rate
to peak at too high a redshift and therefore a direct comparison
between our results and the observational data points can be
misleading. With these caveats in mind, it is nevertheless interesting
to note that the magnitude of the cosmic star formation rate as
derived from our simulations is comparable with that of the
observational data over a considerable range of redshifts.

The overall star formation rate is significantly larger in the
no-feedback case: 
allowing SN to inject energy into the ISM
causes strong galactic winds which 
drive the baryonic gas away
from the halos, while at the same time contributing to the increase of
the temperature and pressure in the ISM, hindering the process of star
formation. As a result, the global star formation rate can take much
larger values in the no-feedback run.

In both runs, there is a change of slope of the cosmic star formation 
rate at around the epoch of re-ionization, 
an effect associated with both reionization and gas depletion. The
radiation field is implemented in the same way in both runs and is
independent of stellar feedback, so reionization is felt equally in
both simulations. However, the decrease of the star formation
activity at low redshifts is much more prominent in the feedback run,
due to the extra effect of gas depletion of halos.
 
\section{Dwarf Galaxies}

\subsection{The Inventory of simulated Dwarf Galaxies}

We find and define the centers of halos using HOP (Eisenstein \& Hut
1998) on the dark matter. A halo is then defined as
the spherical region around these centers which has an average dark
matter density $\Delta_{\rm vir}(z)$ times the critical density
$\rho_{\rm crit}(z)$ (Bryan and Norman 1998).  We require a minimum of 
100 dark matter particles (corresponding to a dark matter mass of 
$\sim 10^9 M_\odot$) in order to identify an overdense region as 
a dark matter halo. Sub-halos (objects identified as halos 
by our algorithm but which are included in larger virialized structures)
have been removed from the resulting set of halos.

\begin{figure}
\plotone{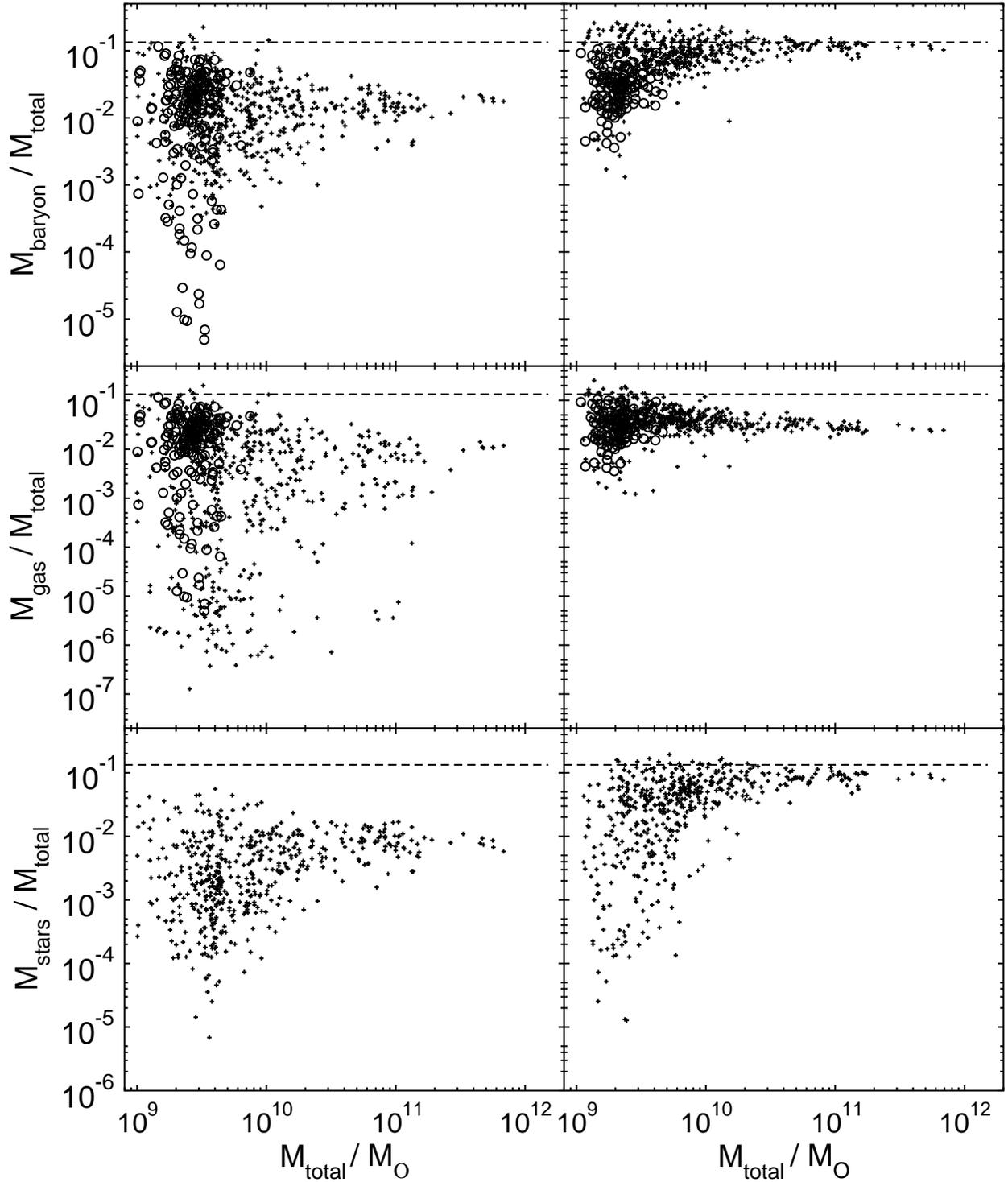}
\caption{
Baryonic content for all galaxies at redshift three. Left panel:
feedback run. Right panel: 
no-feedback run. Top row plots show the baryonic 
ratio for plotted galaxies as a function of their total mass, while 
the second and bottom row plots correspond to their gas and stars 
mass ratio as a function of total mass.
Halos with stars (crosses) and without (open circles) are
shown. The dashed line corresponds to the ratio $\Omega_b/\Omega_m$.
\label{mbmt}}
\end{figure}

Fig.~\ref{mbmt} shows
the variations of the baryonic content (i.e.  stars + gas) of the
simulated galaxies at redshift $z=3$ as well as their gas and stellar
contents vs total mass, for both the feedback run (left panel) and the
no-feedback run (right panel).  Open circles represent halos with no
stellar content while crosses represent halos with stellar content.
The dashed line corresponds to the ratio $\Omega_b/\Omega_m$ (average
ratio of baryon density over total matter density in the universe).

In the no-feedback run, the total baryon content is closely scattered
around the $\Omega_b/\Omega_m$ line, as seen in previous $\Lambda-$CDM
simulations (e.g. Pearce et al. 2001). The deviation from the 
cosmic mean at low masses is due to the effect of photoionization 
on the baryon contents of dark matter halos 
and follows the shape also found by Gnedin (2000).
On the contrary, in the
feedback run, even the largest halos at z=3 consistently contain only
a small fraction of the baryons corresponding to $\Omega_b/\Omega_m$,
which is also suggested by the properties of observed dwarf galaxies
in the local group (e.g. Mateo 1998 and references therein) This
significant difference between the results of the two runs indicates
that low galaxy baryon contents can occur as a result of energy
and momentum feedback from supernovae to the baryon gas: the latter is
driven out of the potential well of the dark matter halos, leaving
behind galaxies of lowered baryon content.

We can draw additional support for this picture by comparing the
two panels in the middle row of figure \ref{mbmt} which displays the 
ratio of gas
mass over total mass for the z=3 halos. In the left panel figure,
which corresponds to the case with stellar feedback, the scatter of
halos towards low gas mass values is much larger than in the
no-feedback case. This fact, considered in the framework of the
scenario described above, is no surprise: in the no-feedback run the
only mechanisms for gas depletion in a halo are the pressure from
photo-ionization (reionization), tidal stripping and the conversion of
baryon gas to stars.  In the simulation where feedback is taken into
account, gas can be lost both to star formation and to the
intergalactic medium because of galactic winds driven by supernovae.
The additional gas depletion mechanism drives galaxies towards lower
gas content values in the feedback run.

By comparing the two bottom panels of figure \ref{mbmt} we can draw
conclusions concerning the stellar population of halos in the feedback
and no-feedback runs. A first obvious result is the existence of a
large scatter in the total stellar content of galaxies both in the
no-feedback and the feedback case.  However, if we compare the
galaxies with the largest stellar contents in both simulations, we
find that the no-feedback run galaxies have much larger stellar
contents, with the ratio of stellar mass over total mass reaching the
$\Omega_b/\Omega_m$ value. This result is consistent with the much
higher value of overall star formation rate in the no-feedback run
when compared to that of the feedback run.  Thus stellar feedback is a
factor that clearly hinders star formation through the depletion of
baryonic gas and the increase of the temperature and pressure of the
ISM.

It is also interesting to note that dark halos which do not host stars
(open circles in Fig. \ref{mbmt}) all have relatively low masses. This
can be understood within a hierarchical clustering scenario in terms
of the inability of such small halos to hold photo-ionized gas.

An additional effect of feedback is that by z=3 there already exists a
population of halos with no stellar content whose baryon gas content
is also extremely low (as low as $10^{-6} \times M_{\rm dark
\,\,matter}$). The observational detection of such halos which are
practically "emptied" from baryonic gas would be extremely
difficult. This result could at least partially explain the deficit of
low-mass satellite galaxies in observations in the Local Group as
compared to the expectations from CDM cosmological simulations (Moore
et al. 1999, Klypin et al. 1999). The difference in mass to light
ratios at small masses caused by reionization has also been
investigated by Somerville (2002) and Benson et
al. (2002). Integration of the effect of stellar feedback to such
simulations naturally drives a portion of the low-mass systems to
baryon contents low enough to render them unobservable by means other
than their dynamical effects.

\begin{figure}
\plotone{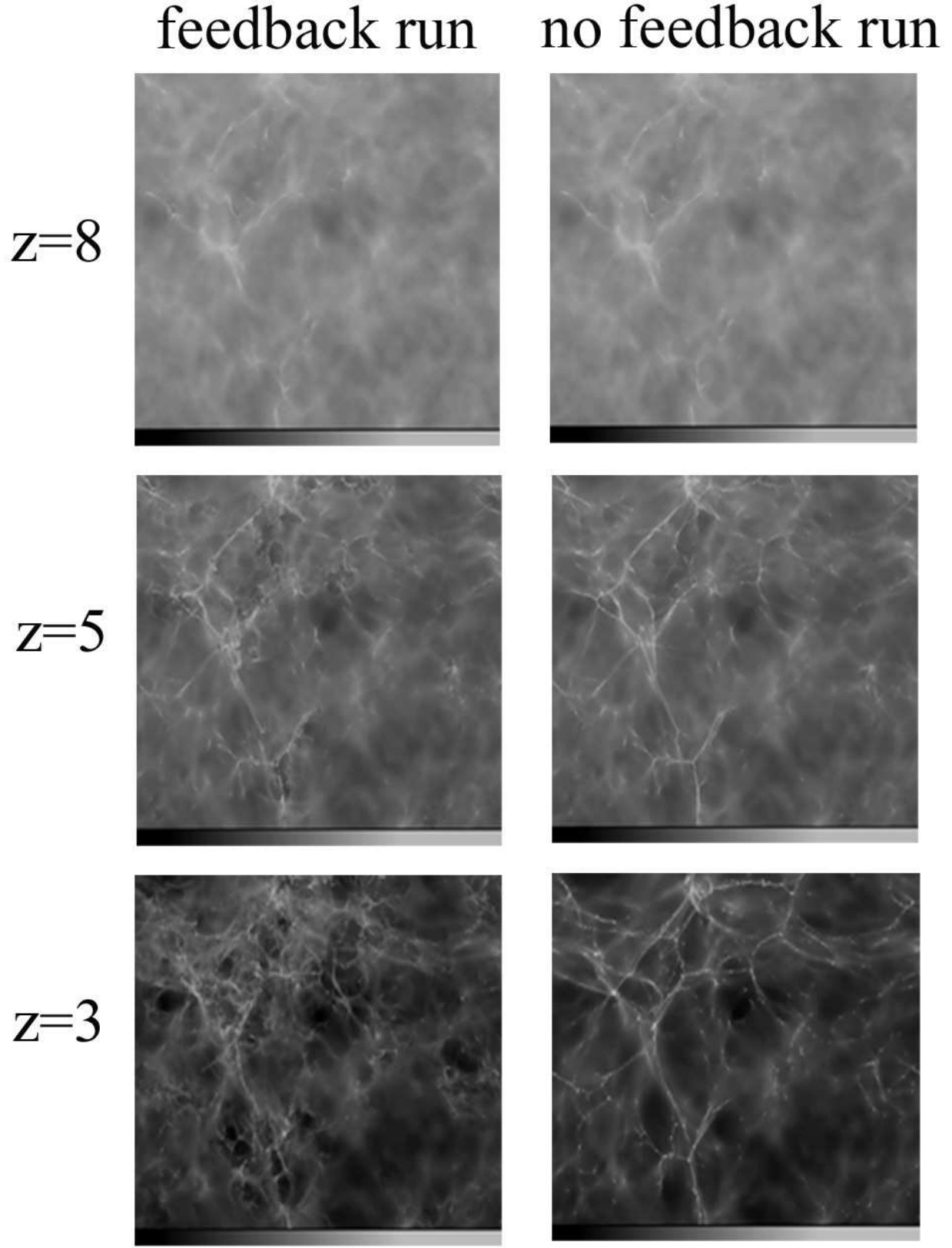}
\caption{Gas density projection along x-axis. 
Range plotted:
$ 11.1 \le \log \frac{\rho _{\rm gas}}{M_\odot / (\rm pr. \,\, 
Mpc^2)}
\le 13.6 $.
\label{fig:feedback1}}
\end{figure}

\begin{figure}
\plotone{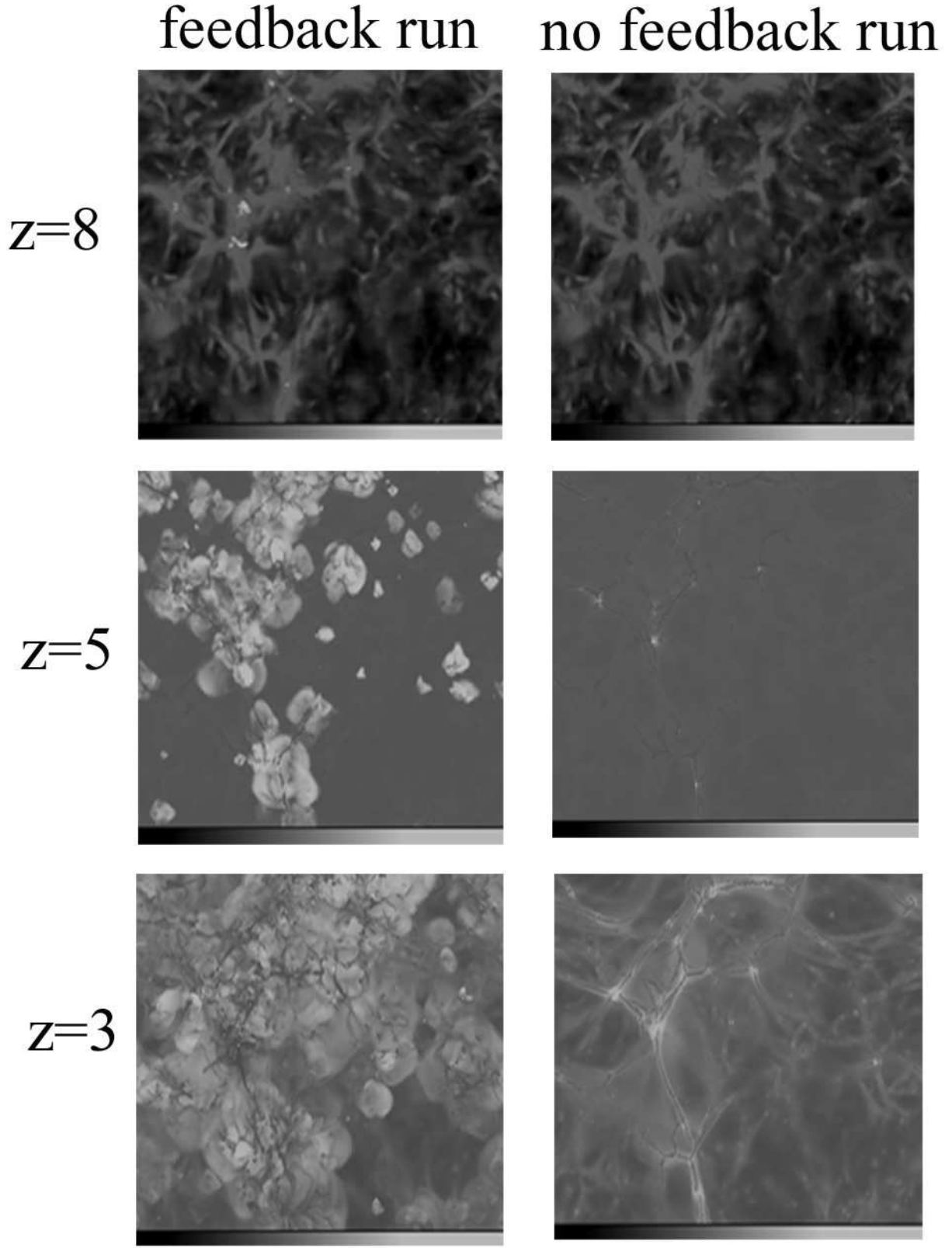}
\caption{X-ray weighted temperature projection along x-axis.
Range:
$  2.0 \le
\log \frac{T}{K}
\le 7.7
$.
\label{fig:feedback2}}
\end{figure}

The morphology and complexity of the gas outflows responsible for 
the modification of the baryon contents of our simulated halos are
demonstrated in figures \ref{fig:feedback1} and \ref{fig:feedback2}. 
In Fig.~\ref{fig:feedback1} the projection along the
x-axis of the gas density in the simulated cube is plotted in a
logarithmic scale for redshifts 8, 5 and 3, and for both the feedback
and the non-feedback runs. The color scale for $\log
[\rho_{\rm gas}/(M_\odot/ {\rm proper \,\, Mpc^2})]$ ranges from
11.1 to 13.6 for all snapshots. The deviation between the feedback
and non-feedback behaviour is evident from $z=5$ and is most striking at
$z=3$. Winds driven by the stellar feedback increase the temperature of
the baryonic gas and drive the gas away from the galaxies in a rather
asymmetrical fashion. 

Fig.~\ref{fig:feedback2}  shows the projection along the x-axis of the x-ray weighted
temperature in our cube, again for redshifts 8, 5 and 3 and for both
the feedback and non-feedback run. The color scale is again
logarithmic, and extends from $10^2$ to about $10^8$ K, and is the 
same for all
snapshots. Feedback is shown to result in the creation of
non-spherical bubbles of hot gas, which, by redshift 3, have a typical
size of about 1 comoving Mpc. The bubbles overlap, resulting in a global
increase of the mean temperature of the gas in the cube, from $\sim 2.2 
\times
10^4$ K in the non-feedback case, to $\sim 5.6 \times 10^4$ K in the
feedback run, for z=3.  The wind-blown bubbles also do noticeably
sweep up material in shells ahead of them. 

\subsection{Star Formation Histories of Dwarf Galaxies at $z\ge3$}

\subsubsection{Galaxies with and without Stellar Content}

\begin{figure}
\plotone{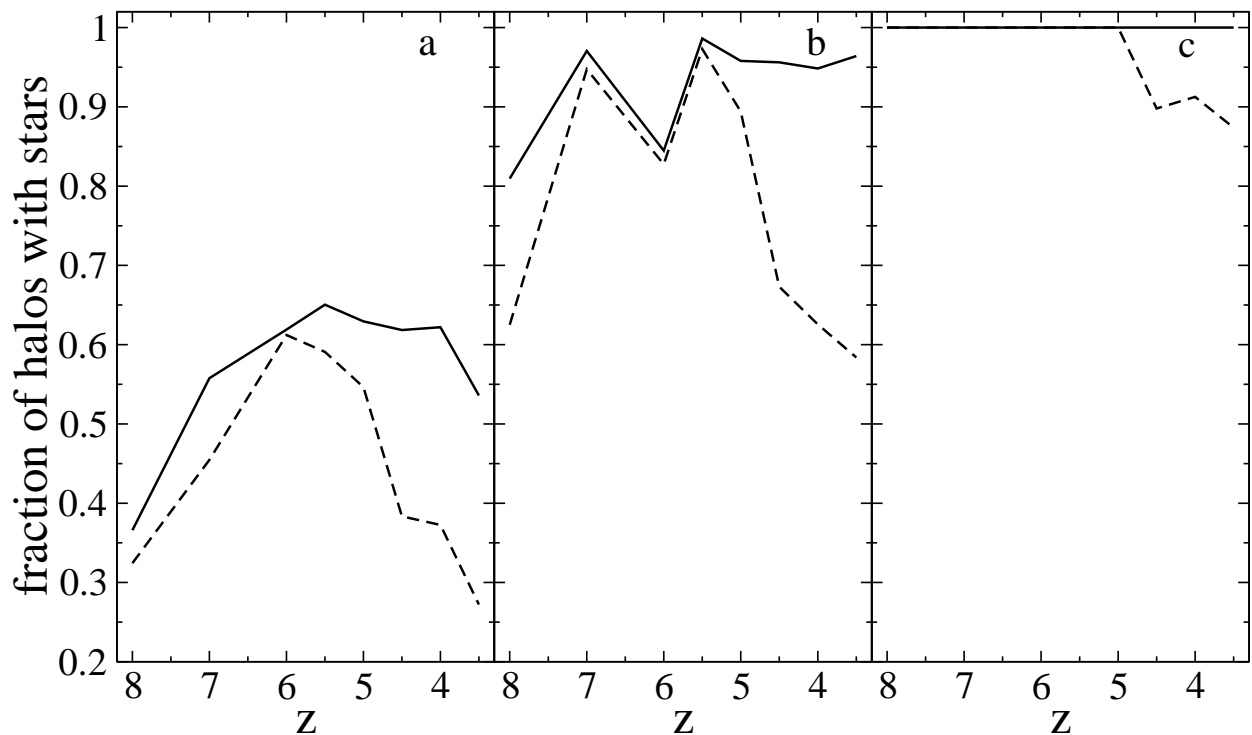}
\caption{Fraction of halos hosting 
stars vs redshift, for 3 
different mass ranges. Panel a: 
$10^9 {\rm M_\odot} \le M_{\rm total}
< 3 \times10^9 {\rm M_\odot}$. Panel b:
$3 \times 10^9 {\rm M_\odot} \le M_{\rm total}
< 6 \times10^9 {\rm M_\odot}$. Panel c:
$6 \times 10^9 {\rm M_\odot} \le M_{\rm total}
< 2\times10^{10} {\rm M_\odot}$. 
Solid line curves correspond to the no-feedback run, 
while dashed line curves correspond to the feedback run.
\label{dwarfPI}}
\end{figure}

The dwarf galaxy stellar content is affected by both feedback 
and photoionization. To demonstrate the relative importance of 
the two effects, we plot, in Fig. \ref{dwarfPI}, the fraction of 
halos with stellar content (i.e. the number of halos which do 
host stars over the total number of halos in this mass range) 
in three different mass ranges 
( between $10^9 {\rm M_\odot}$ and 
$3 \times10^9 {\rm M_\odot}$ for panel a, 
between 
$3 \times 10^9 {\rm M_\odot}$ and  
$6 \times10^9 {\rm M_\odot}$ for panel b and 
between
$6 \times 10^9 {\rm M_\odot}$ and 
$2\times10^{10} {\rm M_\odot}$ for panel c) 
as a function of redshift. The results of the feedback run 
are represented by the dashed curves, while the results of the 
no-feedback run correspond to the solid curves.

Most low-mass halos (panel a) have not yet formed stars by $z=8$ but,
as time progresses, the baryon density and Jeans mass criteria of our
star formation algorithm are satisfied in an increasing number of
these halos and therefore the fraction of halos with stellar
content initially increases.  However, between redshifts 5 and 6, when
photoionization occurs, the process is reversed. The temperature (and
therefore the Jeans mass) increase, cooling rates and typical
densities decrease, and star formation is inhibited.  Halos which
accumulate enough mass to enter this mass range after reionization do
not form stars, and the fraction of halos with stars decreases
(no-feedback run, solid curve). When feedback is added to this
picture, it enhances the effect of reionization, by providing a second
mechanism for increasing the temperature of the interstellar medium
and decreasing the maximum densities reached in a given halo. The
curve maximum occurs at a higher redshift and the fraction of 
halos with stars decreases even more at low z (feedback run, dashed
curve).

In halos of higher virial masses and temperatures, the central baryon
densities are less affected by photo-ionization from reionization
(panels b and c).   
However, in the same mass ranges, feedback is effective in increasing
the ISM temperature and inhibiting star formation and is responsible
for the small decrease in the number of higher-mass halos hosting stars at 
low redshifts, which is 
present only in the feedback run (dashed curves).

\subsubsection{Star Formation Rates of Individual Galaxies}

\begin{figure}
\plotone{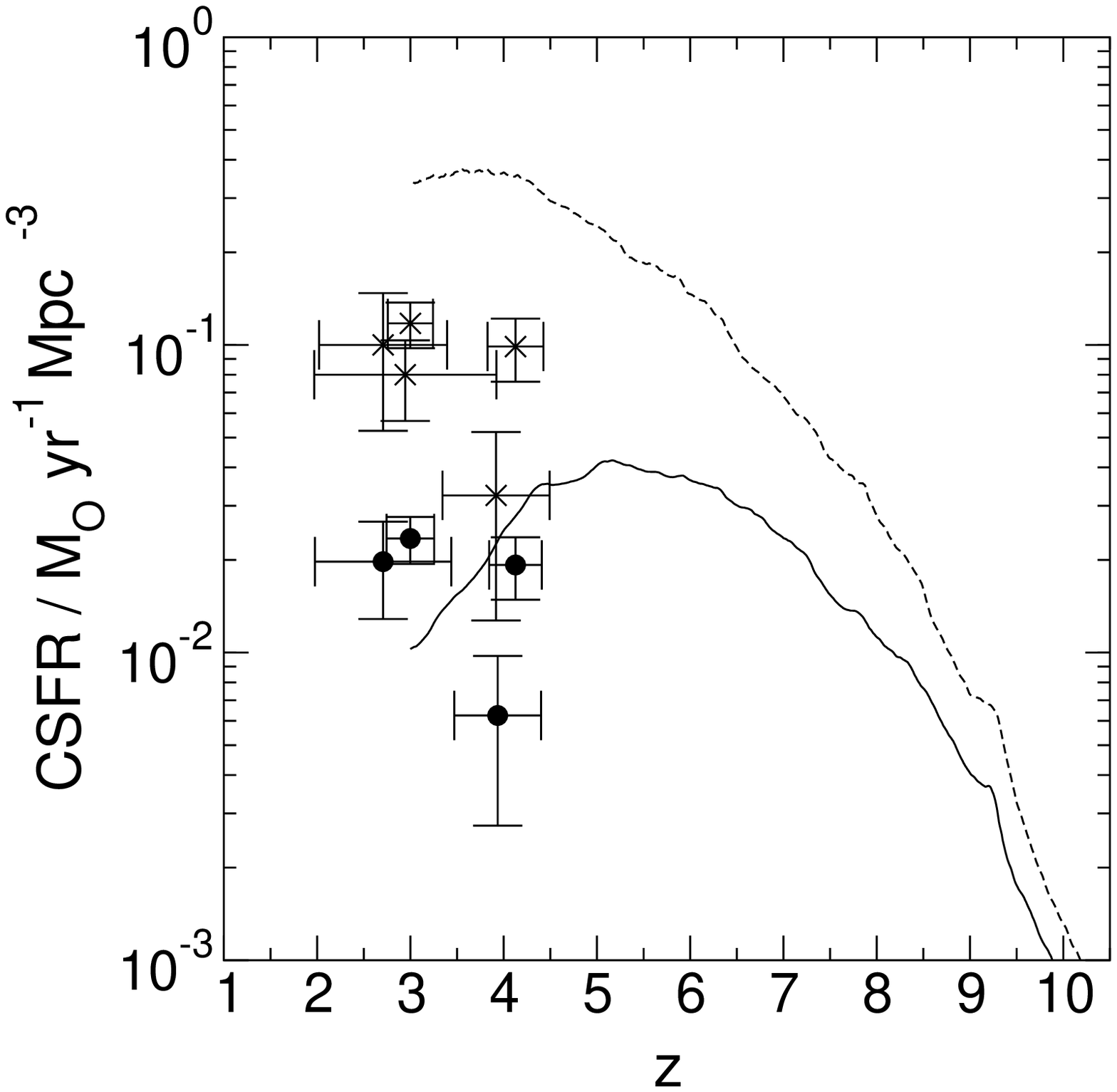}
\caption{The global star formation history. The data points 
were taken from Blain (2001) and references therein. Filled circles
represent measurements uncorrected for dust extinction, while 
diagonal crosses are all measurements which incorporate 
some correction for dust extinction. The solid line
corresponds to the run with stellar feedback and the dashed 
line to the run without feedback.\label{fig:gsfr}}
\end{figure}

In Fig. ~\ref{fig:gsfr} 
we have seen the effect of feedback on the global SFR, which
is to hinder the SF process. We will now examine how the CSFR is
distributed among halos of different SF rates. By extracting
information on the characteristic SFR of the typical SF sites, we can
infer whether the bulk of the SF occurs in sites of large SF activity
which could be observable at high redshifts, or rather in many low-SFR
galaxies which would be unobservable. If the latter is the case, then
observations would tend to underestimate the CSFR at high ($>3$)
redshifts, as has been suggested by Lanzetta \etal (2002).

\begin{figure}
\plotone{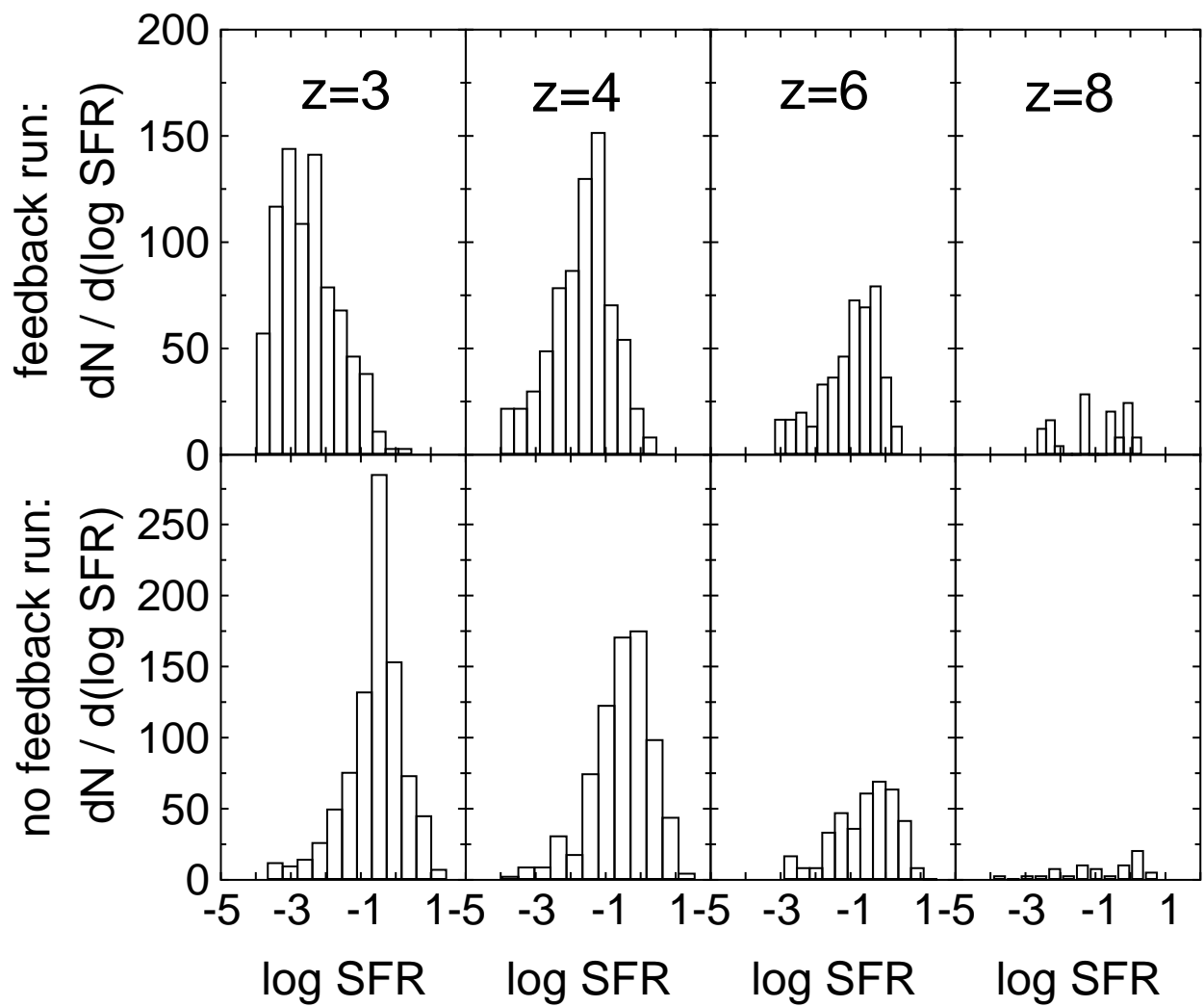}
\caption{Distribution of galaxies per logarithmic
star formation rate interval. Star formation rates are measured in
units of ${\rm M_\odot \, yr^{-1}}$.  The top panel corresponds to the
no-feedback run, while the bottom panel corresponds to the run with stellar
feedback.Results for z=3, 4, 6 and 8 (from left to right) are shown.
\label{sfr_dist}}
\end{figure}

Fig. \ref{sfr_dist} shows the distribution of the number of galaxies
per logarithmic star formation interval for both feedback and
non-feedback simulations and for redshifts z=3, 4, 6 and 8. Star
formation rates are measured in ${\rm M_\odot \, yr^{-1}}$. In the
case with stellar feedback (top panel of Fig. \ref{sfr_dist}), 
as we move  to lower redshifts, the peak of
the distribution moves towards lower star formation values, a result
of the depletion of halos of baryon gas as the latter is being driven
away from the dark matter halo potential wells by stellar explosions.
This effect is absent in the case of the no-feedback run (lower panel 
of Fig. \ref{sfr_dist}). 
Thus, in the
presence of stellar feedback, halos tend to shift towards lower SFR
with advancing cosmic time. By z=3 the distribution of halos in our
simulation per SFR interval peaks around $10^{-3} {\rm \, M_\odot \,
y^{-1}}$.

\begin{figure}
\plotone{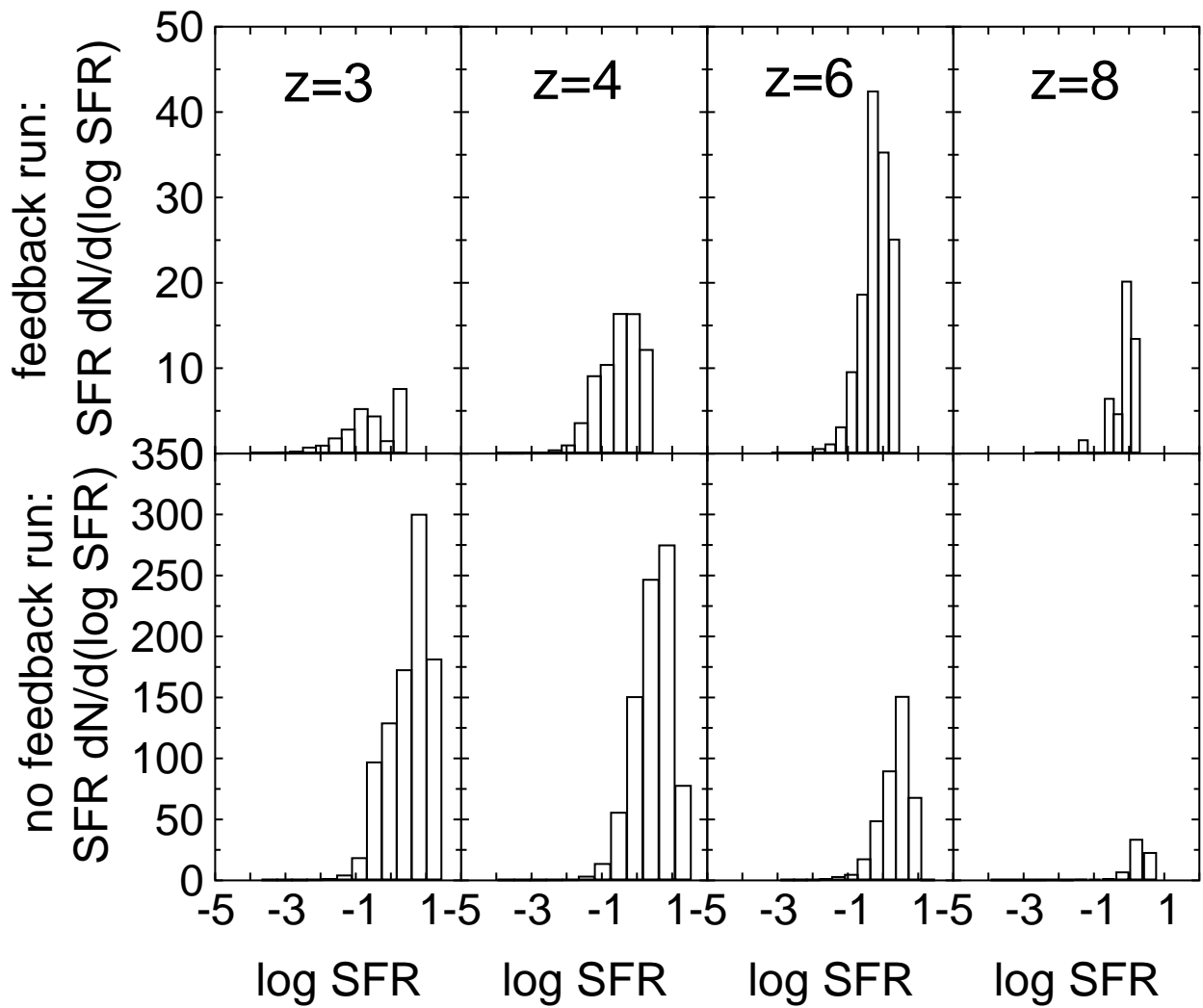}
\caption{
${\rm SFR \times \frac{dN}{d \log SFR} }$ as a function of
${\rm \log SFR}$. The area under the curve plotted is the total
SFR in the simulated cube.\label{xsfr_dist}}
\end{figure}

Although Fig. \ref{sfr_dist} clearly demonstrates that there are many
more halos with low SFR by z=3, this does not necessarily suggest that
most SF activity takes place at low SFR halos. We would like to
determine whether the low SFR halo population has a significant
cumulative contribution to the global SF activity or if the latter is
dominated by the few high-SFR sites.  For this purpose, in
Fig. \ref{xsfr_dist} we plot the distribution of star formation
activity per logarithmic star formation rate interval, $ {\rm SFR
\times (dN/d \log SFR) }$ vs ${\rm \log SFR}$.  The area under this
curve gives the total SFR at the specific redshift due to all star
particles in our cube. Again the plots are for both the feedback and
the non-feedback simulation and for redshifts z=3, 4, 6 and 8. Star
formation rates are measured in ${\rm M_\odot \, yr^{-1}}$.  In the
no-feedback plot, the total SFR saturates after reionization (as also
seen in Fig. ~\ref{fig:gsfr}), while in the feedback case there 
is a significant
decline in the overall star formation rate as we progress to lower
redshifts, due to the additional effect of gas depletion.  It is also
interesting that in the feedback run results, and at redshifts $<5$, a
significant portion of the total star formation is indeed coming from
low-SFR galaxies, potentially unobservable, which might lead to an
underestimate of the cosmic star formation rate from observations of
individual objects after reionization.  However, before reionization,
most SF activity comes from high-SFR halos and starburst-like events.
In addition, when interpreting these results, the reader should keep
in mind that due to the small volume of our simulation, the large
galaxies with high SF are under-represented: On the one hand, our small
volume can only host 4 at most Lyman break galaxies. On the other
hand, the high SF efficiency and the consequent rapid gas depletion
force the SFR to peak at too high a redshift, so that by $z=3$ the
SFRs of our simulated halos are being underestimated.

\subsection{Metallicities of Dwarf Galaxies at z=3}

Prada \& Burkert (2002) have pointed out that there is a correlation
between the mass-to-light ratio M/L of dwarf spheroidal Local Group
satellites and their metallicities, [Fe/H]. They have found that
\[\log (M/L)=-[Fe/H] + {\rm \,\,constant} \,\,.\]

\begin{figure}
\plotone{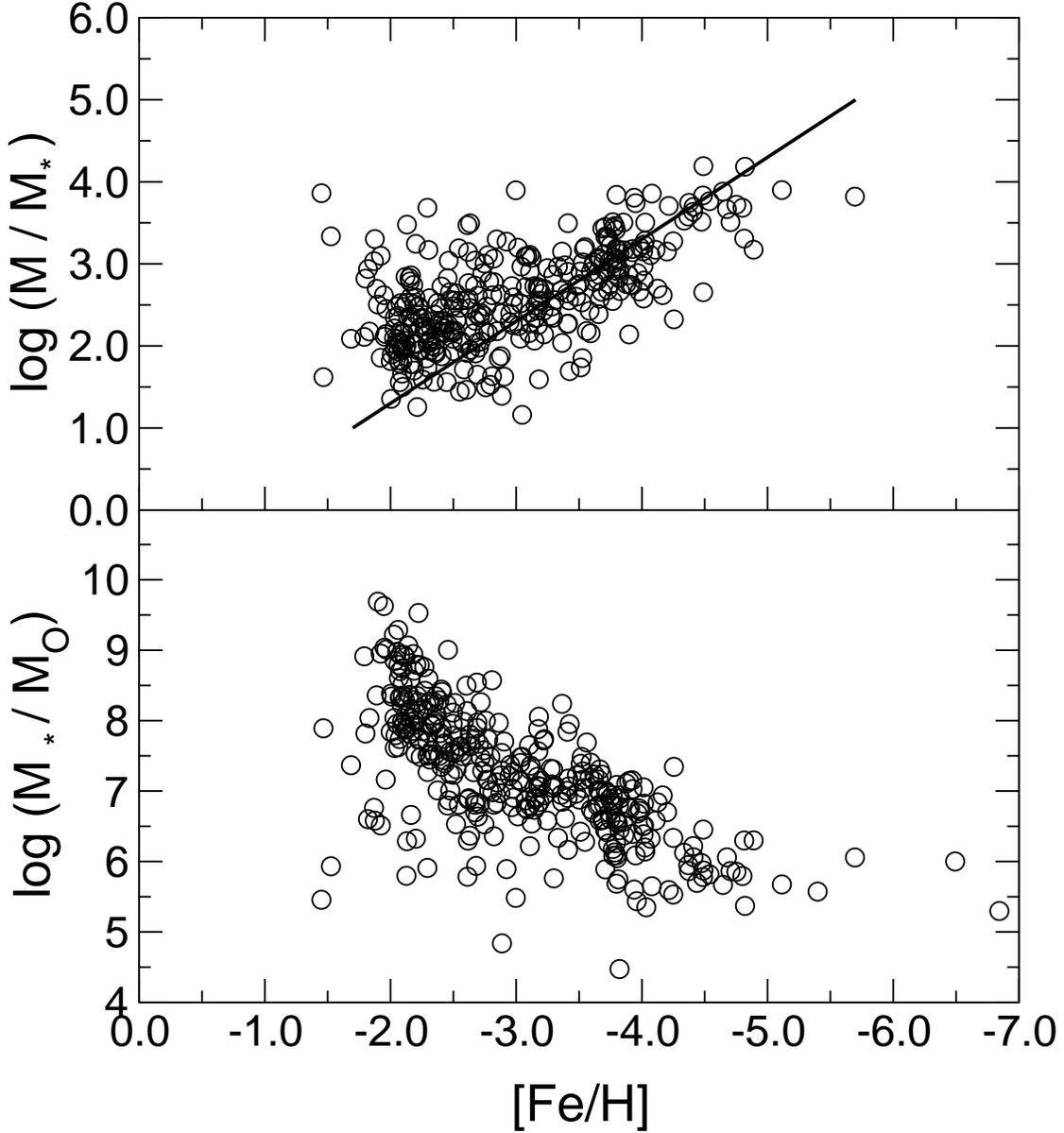}
\caption{Upper panel: 
log($M_{\rm total} / M_{\rm stars}$) vs mass weighted average
metallicity of
stars in each halo.  A clear trend for decreasing metallicity with
increasing $M_{\rm total} / M_{\rm stars}$ is found. The solid line
indicates the slope of the same trend found in observations of dwarf
spheroidal galaxies in the Local Group (Prada \& Burkert 2002).
Lower panel:
stellar masses of halos vs their  metallicities. 
\label{fig:met}}
\end{figure}

In order to look for a similar correlation in the inventory of our
simulated halos, we plot in Fig.~\ref{fig:met} 
the logarithm of the ratio of total
mass over stellar mass ($\log M/M_\star$) vs the metallicity [Fe/H] of
each halo.  To estimate the luminosities of our stellar particles, we
have used $L \propto M_\star$.  This is only an approximation, since
$L$ depends on stellar age and metallicity as well; however, it is
sufficient to estimate the trend of the $M/L$ ratio.  Under the above
assumption, $\log M/M_\star$ only differs from $\log M/L$ by an
additive constant.  In the same plot, we have drawn a line of slope
-1, as in the observational result of Prada \& Burkert (2002), with a 
shifted
intercept (since we are plotting $M/M_\star$ rather than mass-to-light
ratios). Although the scatter of the points is significant, one can
observe the trend for the metallicity to decrease for increasing
$M_{\rm total}/M_\star$.

A large $M/L$ implies a small stellar population relative to the total
mass and consequently less integrated metal enrichment of the ISM than
systems with lower total mass to light ratios. In addition, small 
halos tend to blow out gas and metals and halt star formation before a
high metallicity is reached within them.  Thus, these systems tend to
have lower metal abundances than more massive systems.

When comparing Fig.~\ref{fig:met} with observations, one should keep
in mind that the Local Group satellites are much more evolved objects,
while the evolution of our simulated halos stops at z=3. However, it
is encouraging that the simulations do show this trend and the hope is
that in the future it will be possible to simulate a larger volume at
similar mass and spatial resolution down to redshift zero.  
\section{Discussion}

We have presented the first results of a large-scale three-dimensional
cosmological hydrodynamic AMR simulation of a cube of comoving size
$\sim 10$ Mpc living in a $\Lambda$CDM universe with cosmological
parameters $(\Omega_\Lambda, \Omega_m, h) = (0.7,0.3,0.67)$.  Star
formation was allowed for, according to a parametric prescription
creating collisionless particle out of baryonic fluid condensations
exceeding the local Jeans mass.  We have compared the results of two
runs, differing in the treatment of the stellar particles: in one case
the latter did return energy, momentum and metal feedback to the
interstellar medium, while in the other no stellar feedback was
allowed for. The effects of reionization were studied by subjecting the
baryon fluid to a homogeneous, Haardt \& Madau radiation field of
index -1.5, independent of the cosmic star formation. 

Over 300 galaxies are produced in our simulation over the mass range
$3\times 10^9 \le M_{\rm tot}/M_\sun \le 7 \times 10^{11}$.  The
smallest dwarf galaxies in the local group show kinematic masses of
$2\times 10^7\Ms$ (Mateo 1998 and references therein). However, these
observed galaxies may be hosted in much more extended dark matter
halos (Stoehr et al 2002).  It seems premature to argue for a
significant modification of the cold dark matter scenarios such as to
ask for significant annihilation rates (Cen 2001) to accommodate for
their existence. The 
star formation histories of the smallest dwarf galaxies
are also essential in understanding the recently proposed substructure
problem raised by Moore et al. (1999) and Klypin et al.  (1999).  It
has been argued that the systems associated with the extensive dark
matter substructure found in N--body simulations of galaxy formation
would not be in accord with the observations. This was one of the
reasons for the proposal of Spergel and Steinhardt (2000) that the
dark matter may be self--interacting which has attracted significant
theoretical interest (Burkert 2000; Kochanek and White 2000; Yoshida
\etal 2000, and references therein). The connection between simulated
dark matter substructure and observable quantities is a complex one
and must depend dramatically on the intricacies of star formation and
feedback in the small potential wells of interest (Bullock et
al. 2000, Somerville 2002, Benson et al. 2002).

Our results have shown that feedback from stars is a factor which
can deplete halos of baryonic gas and significantly reduce their 
baryon mass fraction. The gas depletion due to feedback results in
halos having a baryon - to - dark matter mass ratio which is
consistently below the cosmic mean, $\Omega_b / \Omega _{DM}$, even in
the very high mass objects. In addition, feedback can result to the
formation of dark matter halos empty of stars by z=3, and a baryon
fraction so low that their observation would be particularly
difficult. The existence of such a population of dark dwarf halos
could at least partially explain the problem of missing satellites in
the Local Group. 

We have found that feedback and reionization affect the star formation 
rate both globally and locally. The slope of the cosmic star formation 
history curve changes around the epoch of reionization with the effect 
being much more pronounced when feedback is present. In addition, 
most halos assembling after reionization do not form stars by $z=3$, 
an effect which is again enhanced by the presence of feedback.
Furthermore, while before reionization most SF activity originates from
high-SFR objects, after reionization there is a shift towards lower
SFR halos as the source of most cosmic star formation.  Thus, to the
extent that this effect is not an artifact of rapid gas depletion due
to our high SF efficiency, a significant amount of the total star
formation at intermediate redshifts (3-5) might be taking place in
low-SFR halos. This effect has been first discussed by Barkana and
Loeb (2000) and is here reported in full cosmological hydrodynamical
simulations.

Finally, we find that the metallicity of galaxies is correlated with
the mass-to-light ratio, increasing for decreasing metallicity. This
trend is consistent with observational results of dwarf spheroidals in
the Local Group. However, most of the objects in our simulation will
merge to make larger galaxies and only a very small fraction will
survive as isolated dwarf sized galaxies until today. So the presented
comparison again only serves as an illustration  and has as yet little
predictive quality for the formation of dwarf galaxies. Nevertheless,
we find it very encouraging that these first results show that with
yet larger simulations one will be able to make predictions and direct
comparisons with properties of local dwarf galaxies.

\acknowledgements{K.T. would like to thank Brian O'Shea 
and Vasiliki Pavlidou for enlightening discussions. K.T. was partially 
supported by a scholarship from the Greek State Scholarships Foundation. 
T.A. is grateful for the hospitality of the Institute of Astronomy at
Cambridge, UK where this work has been completed.  We thank an anonymous 
referee for comments and suggestions which have improved this paper. 
Simulations were carried out on the Origin2000 system at the NCSA, 
University of Illinois at Urbana-Champaign with support from NSF grant 
AST-9803137 and NRAC allocation MCA98020N.}


\clearpage

\clearpage

\clearpage

\clearpage

\clearpage

\clearpage

\clearpage

\clearpage

\end{document}